\documentclass{aa}
\usepackage{graphicx}
\usepackage{rotating}
\usepackage{afterpage}
\usepackage{amssymb}
\usepackage[sort&compress]{natbib}
\bibpunct{(}{)}{;}{a}{}{,}
\newcommand{\kms}{km\thinspace s$^{-1}$}

\newcommand{\hcn}{H$^{\rm 13}$CN}

\begin{document}
   \title{
   86 GHz SiO maser survey of late-type stars in the Inner Galaxy
   \thanks{Based on observations with ISO, an ESA project with instruments
   funded by ESA member states and with the participations of ISAS and NASA.}
   \fnmsep\thanks{Based on observations carried out with the IRAM 30-m 
   telescope located at Pico Veleta. IRAM is supported by INSU/CNRS (France), 
   MPG (Germany) and IGN (Spain).}
   \fnmsep\thanks{ This is paper no. 16 in a refereed journal based on 
   data from the ISOGAL project.}
   \fnmsep\thanks{The full Tables 2 and 3 are only available in electronic 
   form  at the CDS via anonymous ftp to cdsarc.u-strasbg.fr 
   (130.79.128.5) or via 
   http://cdsweb.u-strasbg.fr/cgi-bin/qcat?J/A+A/(vol)/(page). 
   The complete Figure 7 is available only in  electronic form  at 
   EDP Sciences via http://www.edpsciences.org.}
   }
   \subtitle{I. Observational data}
   \author{M.\ Messineo
          \inst{1}
          \and
          H.\ J.\ Habing
          \inst{1}
          \and
          L.\ O.\ Sjouwerman
          \inst{2}
          \and
          A.\ Omont
          \inst{3}
          \and
          K.\ M.\ Menten
          \inst{4}
          }

   \offprints{M. Messineo}

   \institute{Leiden Observatory, P.O. Box 9513, 2300 RA Leiden, the Netherlands\\
              \email{messineo@strw.leidenuniv.nl; habing@strw.leidenuniv.nl}
         \and
             National Radio Astronomy Observatory, P.O. Box 0, Socorro NM 87801, USA\\
             \email{lsjouwerman@aoc.nrao.edu}
         \and
           Institut d'Astrophysique de Paris, CNRS, 98bis Boulevard Arago, F 75014 Paris, France\\
             \email{omont@iap.fr}
         \and
             Max-Planck-Institut f\"ur Radioastronomie, Auf dem H\"ugel 69, D-53121 Bonn, Germany\\
             \email{kmenten@mpifr-bonn.mpg.de}
             }

   \date{Received June xx, 2002; accepted xxxxxx xx, xxxx
   {\bf Version {\underline \today }}}

  \abstract{
  We present 86 GHz ($v = 1, J = 2 \rightarrow 1$) SiO maser line
  observations with the IRAM 30-m telescope
  of a  sample of 441 late-type stars in the Inner Galaxy ($ -4 \degr <
  l < +30 \degr$).
  These stars were selected on basis of their infrared magnitudes 
  and colours from the ISOGAL and MSX catalogues. 
  SiO maser emission was detected in 271 sources, and their
  line-of-sight velocities indicate that the stars are 
  located in the Inner Galaxy. 
  These new detections double the number 
  of line-of-sight velocities  available from previous SiO and OH
   maser observations
  in the area covered by our survey and are, together with other samples
  of e.g.\ OH/IR stars, useful for kinematic studies of the
  central parts of the Galaxy.
   }

   \titlerunning{86 GHz SiO maser survey  in the Inner Galaxy. I}
   \authorrunning{M.\ Messineo et al.}
   \maketitle
%

\section{Introduction}

There has been a growing interest in studies characterizing the kinematics
and the spatial distribution of stars in the Inner Galaxy ($ 30 \degr <
l < -30 \degr$).  Many  recent studies attempt to determine
 the parameters
that describe the dynamics and structure of the Inner Galaxy, i.e.\ its
central bar and/or its bulge tri-axial mass distribution.  

One approach is to map the
spatial density of a stellar population. This has been done,
e.g., for  stars detected
by IRAS toward the Galactic bulge \citep{nakada91,weinberg92}, 
bulge Mira variables \citep{whitelock92}, bulge red clump stars 
\citep{stanek94} and 
giant stars seen in  fields at symmetric longitudes with respect to the
Galactic centre \citep{unavane98}.

Optical studies of the Inner Galaxy are much hindered by the high
interstellar extinction, which can exceed $A_V \approx 30$
 \citep[e.g.][]{schultheis99}, and thus are limited to
small optical windows \citep{holtzman98,zhao94}.
At infrared and radio wavelengths however,
interstellar extinction is much less severe, or even absent.
Extensive infrared  point source catalogues have recently become available
from the ground based DENIS \citep{epchtein94} and 2MASS \citep{beichman98}
near-infrared (nIR) surveys,  the mid-infrared (mIR)  ISO satellite 
survey  \citep[ISOGAL:][]{omont99,omont02}, 
 and the Midcourse Space Experiment
 \citep[MSX:][]{price97,egan99}. These data have given new
insights into the  {\em spatial} stellar density
distribution in the Inner Galaxy.  
To interpret the information given by the recent observations,
detailed models all include some kind of tri-axiality: a tri-axial
Galactic bulge or  bar
\citep[e.g.][]{debattista02,ortwin02,lopezcorreidora01,lopezcorreidora01b, alard01b}.
However, the bar characteristics such as length,  pattern
speed, and position angle, are still poorly constrained. 

Spatial density studies often  neglect an important measurable 
dimension of phase space: the stellar line-of-sight velocity.
In contrast to the large number of data points in the spatial domain of
phase-space, the available data on the line-of-sight velocities of the stars
is  sparse because it is still   difficult to  measure velocities
from optical or infrared studies.
Asymptotic Giant Branch (AGB) stars with large mass-loss are a valuable
exception, since their envelopes often harbour masers which are strong
 enough to be detected 
throughout the Galaxy and  thereby reveal the
line-of-sight velocity of the star to within a few \kms; frequently
detected maser lines are from OH at 1.6 GHz, H$_2$O at 22 GHz, and SiO at
43 GHz and 86 GHz \citep[for a review see][]{habing96}. 
Previous SiO and OH maser surveys in the Galaxy
have demonstrated that locating the circumstellar masers is an effective 
way to measure line-of-sight velocities of the AGB stars
\citep[e.g.][]{baud79,lindqvist92,blommaert94,sevenster01,
sevenster97a,sevenster97b,sjouwerman98a,izumiura99,deguchi00a,
deguchi00b}.
 
Until recently, only a few hundred stellar line-of-sight velocities
were known toward the inner regions of the Milky Way  ($ 30 \degr <
l < -30 \degr$ and $|b| < 1$), mainly from OH/IR stars,
AGB stars with OH maser emission in
the 1612 MHz line, mostly undetected at visual wavelengths.
This number is too
small to allow for a good quantitative multicomponent
analysis of the Galactic structure and dynamics \citep{vauterin98}.
Obtaining more line-of-sight velocities therefore remains an issue 
of prime importance.
However, masers are rare among stars, because 
 sustaining a maser requires a special physical
environment. 
Most of the mid-infrared brightest OH/IR stars  close to the Galactic 
plane were probably already detected
in the blind OH surveys or in the targeted OH or 43 GHz SiO maser observations 
of colour-selected sources from the 
IRAS survey \citep[e.g.][]{vanderveen88}. 
H$_2$O surveys  \citep[e.g.][]{levine95} are  probably incomplete
because the H$_2$O masers are strongly variable.

SiO maser emission is detected  from several transitions towards
oxygen-rich AGB stars and red supergiants. 
On the basis of  the shape and the amplitude of  their visual light curve 
AGB stars  have been classified as  semi-regular 
(SR) stars  and Mira stars.  
Variable AGB stars may also  be classified as
long period variable (LPV) stars,  when their periods are longer than 
100 days \citep{habing96}.
Almost all OH/IR stars are  variable
and have periods longer than 500 days.
In the IRAS color-color diagram the oxygen-rich AGB stars 
are distributed on a well-defined sequence of 
increasing shell opacity and stellar 
mass-loss rate \citep[e.g.][]{vanderveen88,olnon84}, 
which goes from Miras with the bluest
colors and the 9.7 $\mu$m silicate 
feature in emission, to OH/IR stars with the reddest colors 
and the 9.7 $\mu$m silicate feature in absorption.
The relative strengths of different SiO maser lines are observed 
to vary with AGB type \citep{nyman86,nyman93,bujarrabal96},
indicating that the SiO maser properties depend
on the stellar mass loss rate and  on the stellar variability. 
The ratio of the SiO maser intensities of 43 over 86 GHz is found
to be much lower in Mira stars and in
supergiants than in OH/IR stars. This implies that the 86 GHz ($v = 1$) 
SiO maser transition is a good tool to
measure stellar line-of-sight velocities  of Mira-like stars.
Another advantage is that Mira stars are far more numerous 
than  OH/IR stars.
However, these conclusions are based on small number statistics, 
and have neglected effects of variability.

To significantly enlarge the number of known stellar line-of-sight velocities
we have conducted a targeted survey for the 
86 GHz SiO ($v = 1, J = 2 \rightarrow 1$)
 maser line toward  an infrared selected sample of late-type stars. 
Here we describe the selection of sources
and the observational results. A detailed discussion of the
kinematic and physical properties of the detected stars will be addressed in a
forthcoming paper \citep[][ in preparation]{messineo02b}. 
All velocities in this paper refer to line-of-sight
velocities, measured with respect to the Local Standard of Rest (LSR).

\section{Source selection}

The stars to be searched for maser emission were
selected from a preliminary version of the 
combined ISOGAL-DENIS catalogue
and from the MSX catalogue. The search was limited to the Galactic plane between 
$l = -4\degr $ and $l = +30\degr$ and $|b| \la 1\degr$;
the lower limit in longitude is imposed by the 
northern latitude of the IRAM 30-m telescope.

ISOGAL is a 7 and 15 $\mu$m  survey made with  ISOCAM on board of ISO
 of $\sim$16 deg$^2$,   in selected
fields along the Galactic plane, mostly toward the Galactic centre.
The 7 and 15 $\mu$m observations were generally taken at different 
epochs.
With a sensitivity of 10 mJy (two orders of magnitude deeper than IRAS) 
and a resolution of 3-6\arcsec\ ISOGAL detected over 100,000 objects.  The
combination of the mIR data with the {\it I, J, {\rm and}
K$_S$}-band DENIS photometric catalogue \citep{epchtein94} allows for a
good determination of the nature of these sources. ISOGAL has sampled the
AGB population in the Galactic bulge ranging from the highly obscured, mIR 
luminous OH/IR stars, to the lower mass-loss Mira and SR stars near the tip of
Red Giant Branch, at $K_o \sim$ 8.2 at the adopted 
distance of 8.0 kpc to the
Galactic centre \citep{omont99, glass99, ortiz02,alard01}.
The Midcourse Space Experiment (MSX) is a survey at  five  mIR wavelengths
(from 4.3 $\mu$m [B1 band], to 21.4 $\mu$m [E band]) which covers the
entire Galactic plane to $\pm 5 ^\circ $ Galactic latitude
\citep{egan99}. With a sensitivity of 0.1 Jy in band A (8.28 $\mu$m)
 and a spatial resolution of 18.3\arcsec, the MSX Galactic plane 
survey detected more
than 300,000 objects. 

\begin{figure*}
\includegraphics[width=\textwidth,height=0.4\textwidth]{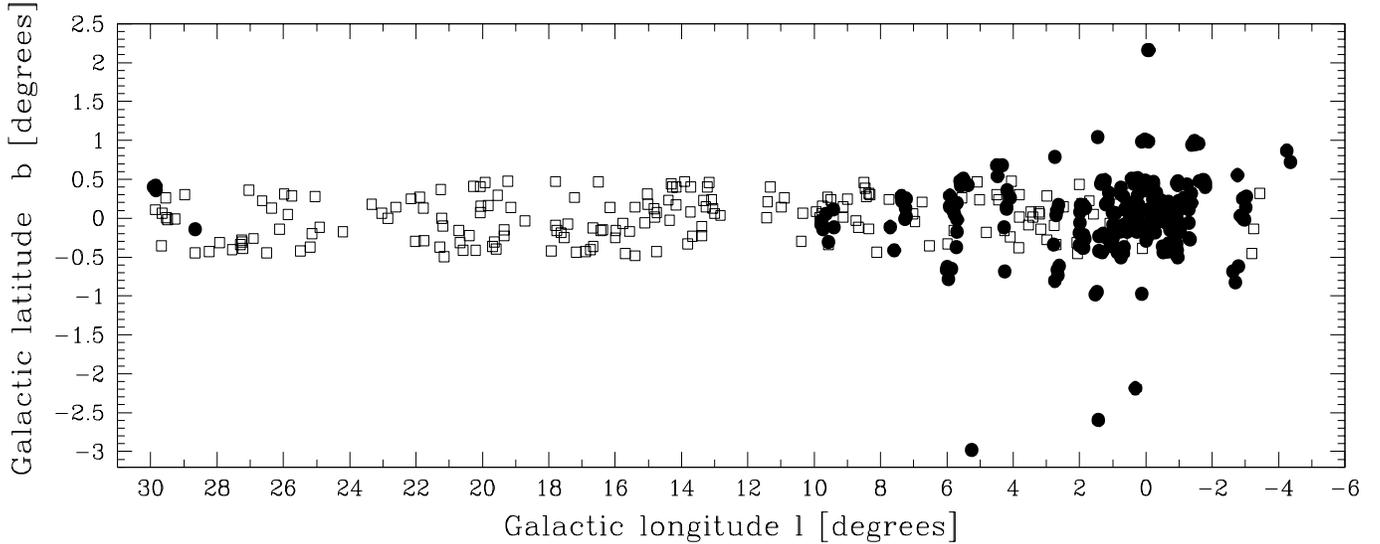}
\caption{\label{fig:lb.ps} 
Location of the observed sources, irrespective of detection and non-detections,
 in Galactic coordinates.
The MSX sources are shown as open squares and the ISOGAL sources 
as filled circles.}
\end{figure*}

Since the ISOGAL survey only covered a limited number of
small fields, the
MSX survey was used to obtain a more even distribution of candidate 
maser sources in the area of interest. Figure \ref{fig:lb.ps} shows
the location of the observed sources.

In the selection of targets, three important earlier observations  
were taken into account as a guideline: 
\begin{itemize}

\item {\it SiO maser emission occurs more frequently towards M-type Mira
stars than towards other AGB stars}
\citep{lewis90,matsuura00,haikala94,bujarrabal94,habing96}.

\item {\it 86 GHz
SiO maser emission is fainter towards OH/IR stars than towards Mira
stars}
\citep[][]{lewis90,nyman86,nyman93}.
\citet{nyman86,nyman93} 
found that the ratio of the 43 over 86 GHz (v = 1) SiO maser line flux 
is  much higher  for OH/IR stars, varying from 2 to 10 with increasing
mass-loss rate,  than for Mira stars (0.4) as found by 
  \citet{lane82}.  
For  typical 43 GHz maser line intensities measured for Galactic bulge 
OH/IR stars \citep{sjouwerman02,lindqvist91}, 
and for the low 86 to 43 GHz  line flux  ratios
generally seen towards OH/IR stars, 
such stars should be weak at 86 GHz and
at the distance of the Galactic centre, 
while Mira stars should be 
stronger.

\item {\it The SiO maser line flux correlates with the mIR continuum
flux density} \citep{bujarrabal87,bujarrabal94,alcolea89,jiang02}.
    Previous studies
have shown that the 43 GHz ($v = 1$ and $v = 2$) line peak intensity  correlates
with the ($\sim$ 7-12 $\mu$m) mIR continuum flux density
\citep{bujarrabal94}.  The mIR and maser intensities vary together
during the stellar cycle.  Measuring the respective flux densities at
the maximum, \citet{bujarrabal87} found that the 
IR$_{8\ \mu{\rm m}}$/SiO$_{43\ {\rm GHz}, v = 1}$ ratio is $\sim$ 5 in 
O-rich Mira stars. 
The IR$_{12\ \mu{\rm m}}$/SiO$_{86\ {\rm GHz}, v = 1}$ ratio 
was found to be $\sim$ 10, albeit with a large scatter, by \citet{bujarrabal96}
in a  sample of O-rich late-type stars.
\end{itemize}

These observations provided us with criteria for our target selection, such as
a lower IR flux density limit in order to detect the maser and  IR 
colours to exclude the high mass-loss AGB stars.  
To avoid OH/IR stars, for which kinematic data is already known, 
we compiled a list of known OH/IR stars by combining the catalogues of
\citet{sevenster97a, sevenster97b, sevenster01},
\citet{sjouwerman98a} and \citet{lindqvist92}.

\subsection{ISOGAL selection criteria}
The ISOGAL catalogue lists the measurements at 7 and 15 $\mu$m 
in  magnitudes, here indicated as [7] and [15], respectively.  
The magnitudes of the sources in the ISOGAL-DENIS catalogue were
corrected for extinction using an extinction map derived from the 
DENIS data by \citet{schultheis99} and using the extinction ratios 
(A$_V$ : A$_I$ : A$_J$ : A$_K$ : A$_7$ : A$_{15}$) 
$ = (1 : 0.469 : 0.256 : 0.089 : 0.020 : 0.025$) 
\citep{glass99,hennebelle01,teyssier02}.

\begin{figure}
\resizebox{\hsize}{!}{\includegraphics{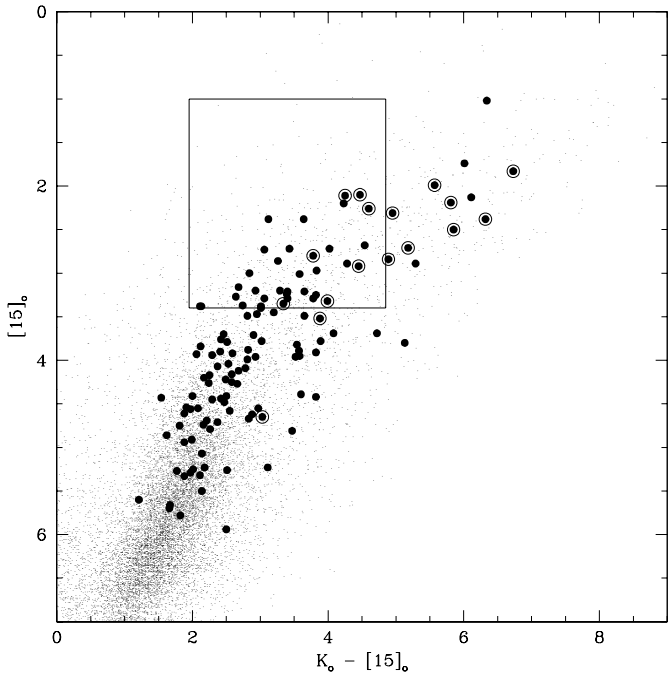}}
\caption{\label{fig:fcsel.ps}  $[15]_o$ versus  ($K_o - [15]_o$)
ISOGAL-DENIS colour-magnitude diagram; extinction correction has been applied.
LPV stars from \cite{glass01} are shown as big dots of which 
the encircled ones mark the variables with known OH maser emission. 
For comparison, the numerous small dots indicate all the ISOGAL sources
within 5 degrees from the Galactic centre. The rectangular
box defines the region of our sample. 
Note that the very red variables without OH emission to the right
of the box  in this figure correspond to outliers without OH emission 
in Fig.\ \ref{fig:fcsel2.ps}.}
\end{figure}

\begin{figure}
\resizebox{\hsize}{!}{\includegraphics{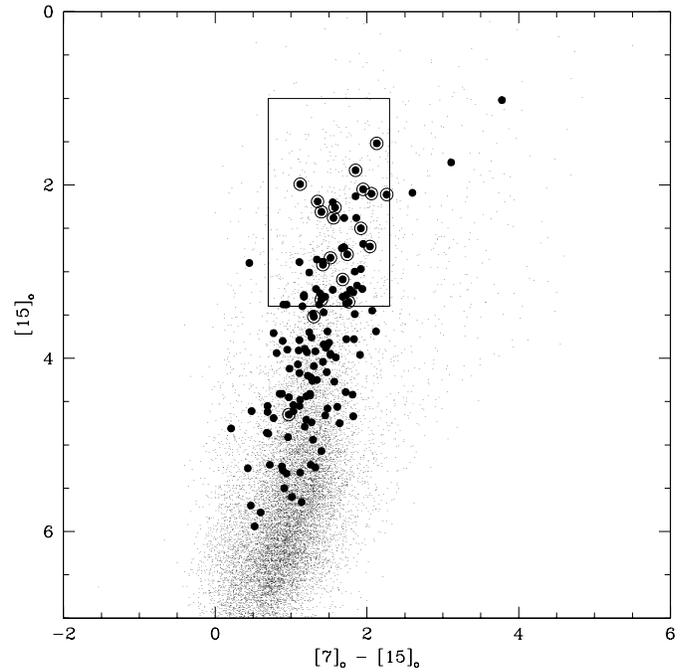}}
\caption{\label{fig:fcsel2.ps}   $[15]_o$ versus ($[7]_o - [15]_o$)
ISOGAL-DENIS colour-magnitude diagram; symbols as in Fig.\ \ref{fig:fcsel.ps}.
Note that  a $K_o$ is not known for all plotted sources, 
and they thus are missing in Fig.\ \ref{fig:fcsel.ps}.}
\end{figure}

We  cross-referenced the Galactic centre LPV positions
\citep[][ 0.5\arcsec\ accuracy]{glass01} with the ISOGAL-DENIS catalogues
and found 180 possible
counterparts out of 194 variables located in fields observed by ISOGAL.
The missing sources can be explained by blending 
with other sources or by high background emission
(the complete description  of this cross-correlation 
will be the subject of a forthcoming paper). Analyzing the
locations of the LPV stars in the different 
ISOGAL-DENIS colour-magnitude diagrams (e.g.\ Fig.\ \ref{fig:fcsel.ps} and
Fig.\ \ref{fig:fcsel2.ps}), where
extinction corrected values are indicated by the suffix "o",
 we found that in the 
$(K_o - [15]_o)$ versus $  [15]_o$ colour-magnitude  diagram the 
LPVs without OH
masers separate well from the OH/IR stars 
(at $(K_o - [15]_o) \approx 4$; Fig.\ \ref{fig:fcsel.ps}).  

The OH/IR stars, having $10^2~ {\rm to}~ 10^3$
times higher mass-loss rates than Mira stars, 
are the brightest objects
at 15 $\mu$m and have   $(K_o - [15]_o)$ colour redder than  4 mag
\citep{ortiz02}. 
This colour is an excellent indicator of infrared emission by the
stellar envelope \citep{omont99}. 

We selected sources from the ISOGAL catalogue by their 
[15]$_o$ magnitude, and their  $(K_o - [15]_o)$ and ($[7]_o-[15]_o$) colours.
See the search boxes in Fig.\ \ref{fig:fcsel.ps} and  \ref{fig:fcsel2.ps}.
We excluded the brightest 15 $\mu$m
sources, [15]$_o < 1.0$, and those with ([7]$_o-[15]_o) <  0.7$ and with
($K_o-[15]_o) <  1.95$, since they are likely to be foreground stars. We
further excluded sources with $[15]_o >$ 3.4 since they are --given the
general correlation of SiO maser emission and IR luminosity-- likely
to show SiO maser emission fainter than our 
detection limit of 0.2 Jy.  Sources with ($[7]_o-[15]_o) > 2.3 $ were excluded since
they are likely to be compact H\ion{II}\ regions or young stellar objects
\citep{felli00,schuller02}, and those with ($K_o-[15]_o) > 4.85$ 
because they are likely to be  OH/IR stars with a  high mass-loss rate 
(Fig.\ \ref{fig:fcsel.ps}) or young stellar objects.
To conservatively avoid duplicating the 
OH maser line-of-sight data points,
sources near (50\arcsec) a known OH maser were excluded. 

As the final photometry of ISO has changed slightly from the
preliminary input catalogue, 16 of the selected sources no longer 
 obey the selection criteria strictly.
253 objects were observed from the selected ISOGAL-DENIS sources.

\subsection{MSX selection criteria}
Since the MSX catalogue gives the source flux density, $F$, in Jy, 
here we use this unit. Magnitudes are obtained 
adopting as  zero points: 58.55 Jy in A (8.26 $\mu$m) band,
26.51 Jy in  C (12.12 $\mu$m) band, 18.29 Jy in D (14.65 $\mu$m) band 
and 8.75 Jy in E (21.41 $\mu$m) band. 

NIR data from DENIS or 2MASS was not available for the MSX sources 
 at the time of our observations.
For the source selection  we used  flux densities  in the A and 
D bands which have  wavelength ranges similar to 
 the ISOGAL 7 and 15 $\mu$m bands.
We selected those not-confused, good quality sources in band A and D
(flag $> 3$), which show variability in band A.  We avoided the
reddest, $F_D/F_A >2.29$, sources, which are likely to be OH/IR
stars or young stellar objects, and the bluest and most luminous
(likely foreground) stars with $F_D/F_A< 0.63$ and $F_D >6$ Jy.
Furthermore, following the work of \citet{kwok97} on IRAS sources with
low-resolution spectra, we used the C  to E 
band ratio to discard very red ($F_E/F_C > 1.4$) sources, 
which are likely to be young stellar objects or OH/IR stars with 
thick envelopes.
Moreover, sources within 50 arcsec of a  known OH maser 
were discarded, as the kinematic data are already known.
We observed 188 sources from this MSX-selected sample.

\section{Observations and data reduction}

The observations were carried out with the IRAM 30-m telescope (Pico
Veleta, Spain) between August 2000 and September 2001 (Table
\ref{table:obs}).  Two receivers were used to observe the two orthogonal linear
polarizations of the SiO ($v = 1, J = 2 \rightarrow 1$) transition at
86.24335 GHz. For each receiver we used one quarter of the low resolution, 1 MHz
1024 channel analog filter bank (3.5 \kms\ spectral resolution, and 890 \kms\ total
velocity coverage),  and in parallel  the AOS autocorrelator
at a resolution of 312.5 kHz (1.1 \kms) with a bandwidth of 280 MHz
(973 \kms\ total velocity coverage).
The telescope pointing errors were typically 2-4\arcsec,
which is small compared to the beam FWHM of 29\arcsec.  
The observations
were made in wobbler switching mode, with the wobbler throw varying
 between 100 and 200 arcsec.
The on-source integration time was between 5 and
20 minutes per source, depending on the system temperature
 which varied between 100 and
300 K because of the weather, 
source elevation (typically 10 to 30\degr), 
and amount of continuum emission in the beam. 

Flux calibration was done in a standard way from regular
observations of a hot (ambient) and cold (liquid nitrogen) load. 
A sky-opacity-correction was computed from measuring the blank sky
emissivity and using a model of the atmosphere structure. The
conversion factor from antenna temperature to flux density changed
from 6.0 to 6.2 Jy K$^{-1}$ on  December 12th 2000.

\begin{table}
\centering
\caption{\label{table:obs} IRAM 30-m observing dates}
\begin{tabular}{|l l l|}
\hline
Period No.& Dates & JD-2450000\\
\hline
1 & 26-27 August 2000 & 1782-1783\\
2 & 04-17 December 2000 & 1882-1895\\
3 & 22-24 January 2001 & 1931-1933\\
4 & 09-24 February 2001 & 1949-1964 \\
5 & 23-28 May 2001 & 2052-2057\\
6 & 15 August - 04 September 2001 & 2136-2156 \\
\hline
\end{tabular}
\end{table}

\subsection{Flux stability}
\begin{figure}
\resizebox{\hsize}{!}{\includegraphics{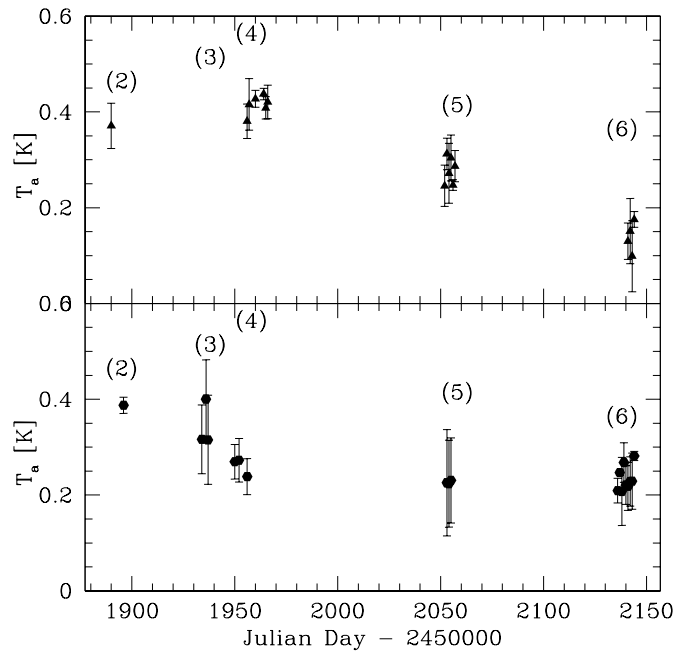}}
\caption{\label{fig:soli.ps} 
Antenna temperatures versus time for the 2 'reference' lines: 
sources  \#265 
plotted with filled hexagons (bottom) and 
 \#163 
with triangles (top).
The observing date on the horizontal axis is expressed in Julian day 
minus 2450000.
The labels refer to the observing periods in Table \ref{table:obs}. 
The measurements of a single period agree within 20 \%.
A long term intrinsic variation (up to a factor of 2 in T$_{\rm a}$) 
is apparent in both sources. }
\end{figure}

Two of the strongest SiO maser sources that were found
in December 2000, were subsequently monitored in order to test
 them as possibly secondary flux calibrators.
Figure \ref{fig:soli.ps}
shows the measured source fluxes, in terms of antenna temperature,
as a function of time over nine
months. Within each observing period (up to a few
weeks) the line flux  
measurements of the two sources are consistent within
the measurement errors, with a typical day-to-day  flux variation of 20 \%.
Over the whole period, however, the flux of the two sources
varied up to a factor two.
We did not notice variations of the average system
temperature that might have caused systematic errors of this magnitude
 in the long-term flux calibration.
The flux  variation of both sources 
is therefore due to intrinsic source variability, 
and this makes these sources of limited use 
as long-term secondary  flux calibrators.  
We shall adopt the observed short-term apparent flux  variations as
an indication of the absolute flux  uncertainty 
for all our observations (i.e.\ $\la 20$ \%).

\subsection{Detection criteria}

The data were reduced using the CLASS software package.  The spectra
taken with the two receivers were combined, yielding a typical rms of
15 mK ($\sim$ 100 mJy) in the AOS channels.  The
line width (FWHM) and the integrated antenna temperature were
determined by fitting the data from the autocorrelator with a gaussian
after subtracting a linear baseline.
Bad channels were eliminated by comparing the analog filter bank and 
autocorrelator spectra.

We considered as a detection
only lines with peak antenna temperature greater than three times the
rms noise level in the autocorrelator spectrum at the original
resolution.
Because of possible  confusion with \hcn\  lines, a problem discussed in  the
following subsection, single component emission lines  detected at
line-of-sight velocity less than $-$30 \kms\ were interpreted as SiO maser lines only if their line width is smaller than 7.5 \kms.

\subsection{Confusion with interstellar H$^{\sf 13}$CN  emission}

The total autocorrelator spectral bandwidth is 280 MHz, centered at the 86243.350 MHz rest
frequency of the ($v = 1, J = 2 \rightarrow 1$) SiO maser transition.
This observing band also includes  the three ($J = 1 \rightarrow 0$) 
hyperfine transitions
of \hcn\ at 86338.767, 86340.184, and 86342.274 MHz.
The \hcn\  lines show up at velocity offsets of
about $-$335 \kms\ ($-$329, $-$336, $-$343 \kms)
relative to the SiO line.

\begin{figure}[h]
\resizebox{\hsize}{!}{\includegraphics{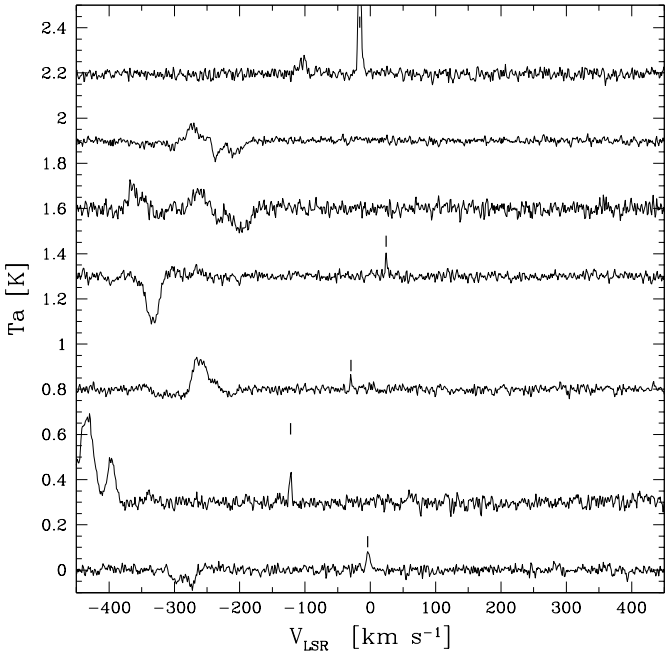}}
\caption{\label{fig:hcnex.ps} Some spectra with interstellar \hcn\ 
emission, shown for clarity with arbitrary antenna temperature offsets.
The small thick vertical bars indicate the detected SiO
maser line. The \hcn\  and SiO lines are usually clearly distinct.}
\end{figure}

The \hcn\ line was observed in many interstellar clouds
and also in the direction of Sgr A \citep[e.g.][]{hirota98,fukui77}.
In our spectra this \hcn\ line was also detected in the direction of
the Galactic centre, at $3.5^\circ<l<-1.5^\circ, |b|<0.5^\circ$, 
in 55 \% of our pointings.
\hcn\ spectra generally  have multiple broad components and
appear in absorption as well as in emission,
 depending on the line intensity in the on- and off-target position 
 (Fig.\ \ref{fig:hcnex.ps}).
The ($l, b$) distribution of the
spectra that contain the \hcn\ ($J= 1 \rightarrow 0$) line
 is similar to that of the Galactic centre molecular
clouds, confirming that the origin is interstellar
\citep[see for example the $^{13}$CO distribution in Fig.\ 2
of][]{bally88}. The \hcn\ line has  been detected also  from
circumstellar  envelopes of carbon stars \citep[e.g.][]{dayal95}, 
but will not be detectable  in AGB stars at 
the distance of the Galactic centre \citep[e.g.][]{olofsson98}.

\begin{figure}[h]
\resizebox{\hsize}{!}{\includegraphics{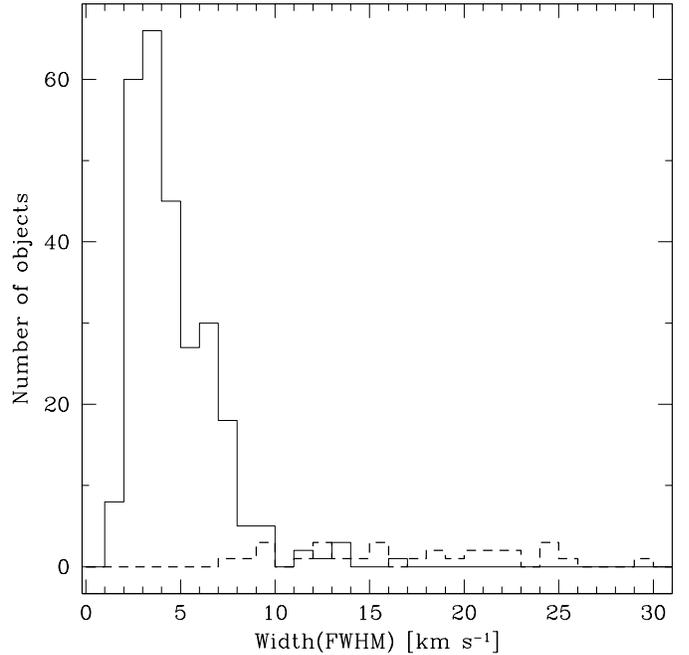}}
\caption{\label{fig:wid.ps} Histogram of the line widths.
The solid line shows the distribution of the SiO line widths; 
the dashed line shows the distribution of the lines with a single component 
in emission which we classified as  \hcn\ emission.}
\end{figure}

Considering that the maximum gas velocity observed in the Galactic
centre is  less than 300 \kms, for example in the CO distribution \citep[e.g.][]{dame01}, 
and considering the frequency differences between 
the three \hcn\ ($J= 1 \rightarrow 0$) hyperfine transitions and the 
SiO ($v = 1, J = 2 \rightarrow 1$) transition, 
the \hcn\ ($J= 1 \rightarrow 0$) line may be confused with  SiO
($v = 1, J = 2 \rightarrow 1$) lines  at velocities 
below $-$30 \kms. All lines detected  at those velocities are
therefore  suspect  and their line  widths were examined in 
order to distinguish between SiO maser and \hcn\ lines.
The typical line width of the \hcn\ and SiO emission is very
different (Fig.\ \ref{fig:hcnex.ps}). 
From the SiO lines detected at velocities 
larger than  $-$30 \kms, i.e.\ where no confusion is
expected, the SiO line width distribution 
ranges between 1.7 and 16.3 \kms\ with a  peak at 
$\approx$ 4 \kms\ (Fig.\ \ref{fig:wid.ps});
the \hcn\ line is much wider and can be up to 100 \kms\ wide.  
With the spectral resolution of the AOS, 1 \kms, in case the 
\hcn\ emission is not spatially extended, 
one  should be able to resolve two or three of the hyperfine components,
 which are separated by 7 \kms.
We therefore identify a spectral line at velocity 
below $-$30 \kms\
as an SiO maser only if it is a single emission component and if its line
width (FWHM) is narrower than 7.5 \kms.  For an unambiguous SiO
identification further observations would be required, 
e.g.\ by searching for the 43 GHz
SiO maser lines, or by using interferometric 86 GHz observations 
to locate the position and 
to determine the extent of the emission.

We have probably rejected  a few SiO lines because of suspected confusion
 with \hcn. 
Of the 202 SiO lines detected at velocities larger than $-30$ \kms\
25 (12 \%)  have  widths larger than 7.5 \kms.
Within 3$^\circ$ from the Galactic centre, we found 75 SiO lines
at velocities larger than $-30$ \kms; while 
at velocities smaller than $-30$ \kms\ 
we found  51  emission line sources with widths $< 7.5$ \kms\ (SiO lines)
and 29 emission lines  with widths $> 7.5$ \kms\ and with single component
which we conservatively classified as \hcn. 
Considering the above 12 \%,
we estimate that we could have eliminated about 7 real SiO lines.

Only one spectrum taken at a position   
away from the central molecular complex
shows a line which is classified as \hcn\  emission 
(at $l = 23 ^\circ$ for source \#423); this is further discussed 
in Sect.\ 4.6.

\subsection{Confusion with interstellar SiO  emission}
In the interstellar medium almost all Si atoms are
supposedly locked up in silicate dust particles.
There is, however, some 
SiO emission  seen  towards interstellar molecular clouds
\citep{martinpintado92,martinpintado97,lis01},  in which case
shocks have probably destroyed the refractory grain cores,
releasing the silicon  back into the gas phase.

However, our stellar SiO maser survey is little
affected by confusion with this interstellar SiO
emission for several reasons.
First, we have targeted mIR sources. The bulk of the
interstellar SiO emission in the Galactic centre is not associated
with mIR radiation \citep{martinpintado97}. Furthermore, our SiO maser
lines are generally narrower than the 10-50 \kms\ found for shocked and
energetic outflows associated with young stars.
Young stellar objects would also be located in a different region
of the colour-magnitude diagrams of Figs.\  \ref{fig:fcsel.ps} and
 \ref{fig:fcsel2.ps} \citep{felli00}.
Finally, follow-up observations with the
Very Large Array (VLA) at 43 GHz have shown that 38 of
39 sources for which we  detected  86 GHz SiO emission
do also show unresolved 43 GHz SiO emission \citep[][ in preparation]{sjouwerman02b}.
It is therefore unlikely that any of this emission arises from
interstellar molecular clouds.

We have also analyzed the 7 SiO lines 
(from Table \ref{table:detections}) which have  widths larger
than 10 \kms\ (\#113, \#117, \#129, \#135, \#173, \#203, \#223). They
represent 2.5 \% of our total number of detections and are located at  longitudes 
between 2 and 18 degrees.
The corresponding seven targeted mIR objects were all detected 
in the DENIS $K$
band and some  in the $J$ band, and their IR colours are typical of
late-type stars. 
Towards two of the sources, \#113 and \#117, we also detected \hcn\ emission
with a difference in radial velocity between the \hcn\ and the 86 GHz
SiO line of $-$42 and $-$22 \kms, respectively. 
The fit of the width of \#113 is however noisy and can very well be less
than 10 \kms\ or a blend of two lines.
For \#117 it could be a molecular cloud line, since a difference 
in radial velocity
between the \hcn\ and the SiO lines up to 25 \kms\ has been already
observed towards molecular outflows \citep{martinpintado92}. However, \#117 is
clearly a $\la$ 7 \kms\ wide, 43 GHz ($v = 1$ and $v = 2, J = 1 \rightarrow 0$) 
SiO maser point source in our VLA follow-up observations \citep[][ in preparation]{sjouwerman02b}.
We conclude that the mIR emission, and 43 and 86 GHz SiO lines are all 
related to an AGB star, although we cannot completely rule out interstellar 
\hcn\ emission for source \#117.

Two SiO emission features are probably of interstellar origin;
see comments in Sec.\ 4.6 for remarks on the individual sources \#94 and \#288.

\section{Results}

\begin{table*}
\caption{\label{table:detections} Sources with SiO maser detected$^{\mathrm{*,**}}$.
The conversion factor from antenna temperature to flux density is 
6.2 Jy K$^{-1}$. }
\medskip  
\ \\ 
$$ 
\begin{array}{cccrccccrll} 
\noalign{\smallskip}  
\noalign{\hrule} 
\noalign{\smallskip}  
 {\rm ID} & {\rm RA }     & {\rm DEC}    &{\rm V}$$_{\rm LRS}$$ &	{\rm T} $$_{\rm a}$$ & & {\rm rms}  &    {\rm A}		&    {\rm FWHM}& {\rm Obs. Date} &{\rm Comments} \\ 
     &  [{\rm J}2000]&  [{\rm J}2000] & [{\rm km\ s}$$^{-1}$$] & [{\rm K}] & &[{\rm K}]  & [{\rm K\ km\ s}$$^{-1}$$] &[{\rm km\ s}$$^{-1}$$]	 & [{\rm yymmdd}]  & \\ 
\noalign{\smallskip}  
\noalign{\hrule} 
  1   & 17\:31\:40.9  & $$-$$32\:03\:56  &      $$-$$8.3 &   0.120 &   &    0.018 &$$   0.34\pm  0.05$$&$$    2.7\pm   0.5$$&       010216 &              \\	   
  2   & 17\:36\:42.2  & $$-$$30\:59\:12  &    $$-$$146.2 &   0.089 &   &    0.023 &$$   0.26\pm  0.08$$&$$    3.0\pm   1.4$$&       010219 &              \\	   
  3   & 17\:37\:07.3  & $$-$$31\:21\:31  &    $$-$$209.2 &   0.079 &   &    0.017 &$$   0.26\pm  0.05$$&$$    3.5\pm   0.9$$&       010216 &              \\	   
  4   & 17\:37\:29.4  & $$-$$31\:17\:17  &    $$-$$232.0 &   0.110 &   &    0.023 &$$   0.52\pm  0.10$$&$$    6.4\pm   1.5$$&       010219 &              \\	   
  5   & 17\:38\:11.5  & $$-$$31\:46\:30  &      10.2 &   0.059 &   &    0.010 &$$   0.20\pm  0.03$$&$$    3.5\pm   0.6$$& 010524/010528 &              \\	   
  6   & 17\:38\:12.5  & $$-$$29\:39\:38  &     113.4 &   0.074 &   &    0.017 &$$   0.48\pm  0.07$$&$$    7.9\pm   1.3$$&       010526 &              \\	   
  7   & 17\:38\:17.0  & $$-$$29\:42\:32  &      18.9 &   0.060 &   &    0.015 &$$   0.36\pm  0.10$$&$$    8.3\pm   4.1$$&       010822 &              \\	   
  8   & 17\:38\:29.0  & $$-$$31\:26\:16  &     $$-$$38.7 &   0.039 &   &    0.011 &$$   0.22\pm  0.04$$&$$    6.3\pm   1.1$$& 010219/010903 &              \\	   
  9   & 17\:38\:32.5  & $$-$$31\:20\:43  &     $$-$$73.4 &   0.082 &   &    0.013 &$$   0.45\pm  0.05$$&$$    6.4\pm   0.8$$&       000827 & 29716$$^{\mathrm{b}}$$\\
  10   & 17\:38\:35.7  & $$-$$29\:36\:38  &       1.5 &   0.074 &   &    0.018 &$$   0.21\pm  0.04$$&$$    2.2\pm   0.4$$&       010523 &              \\	   
\noalign{\smallskip}  
\noalign{\hrule} 
\end{array} 
$$ 
\begin{list}{}{}
\item[$^{\mathrm{*}}$]
The full table contains 271 objects and is available in electronic 
form at the CDS via anonymous ftp to cdsarc.u-strasbg.fr (130.79.128.5)
or via http://cdsweb.u-strasbg.fr/cgi-bin/qcat?J/A+A/(vol)/(page).
\item[$^{\mathrm{**}}$] 
Question marks denote marginal detections (T$_a$/rms $<$ 3.5 ).
\item[$^{\mathrm{a}}$]
Identification number from Table 2 of \citet{glass01}. 
\item[$^{\mathrm{b}}$]
Identification number from Table 2 of \citet{schultheis00}.
\end{list}
\end{table*}

Tables \ref{table:detections} and \ref{table: non-detections} summarize, 
in order of RA, 
our 271 SiO maser detections and 173 non-detections, respectively. 
The columns of Table \ref{table:detections} are as follows: 
an identification number ($ID$), followed by the Right Ascension ($RA$), 
and Declination ($DEC$),
(in J2000) of the telescope pointing, the velocity
of the peak intensity ($V_{\rm LSR}$), as well as the peak antenna temperature
 ($T_{\rm a}$)  and the rms noise ($rms$), the integrated flux density 
($A$) plus its formal error and the line
width ($FWHM$) with its formal error, and finally the observing date ($Obs.Date$). 
If appropriate, comments are added in 
an extra column. The line width was calculated using  a gaussian fit.
Table \ref{table: non-detections}, with the non-detections,
 lists only an identification number ($ID$), 
the  $RA$ and $DEC$ of the telescope pointing, the achieved noise ($rms$) and the observing date ($Obs.Date$). 
An additional column is used for comments on individual pointings.

\begin{figure*}
\centering
  \includegraphics[width=\textheight,height=\textwidth,angle=90]{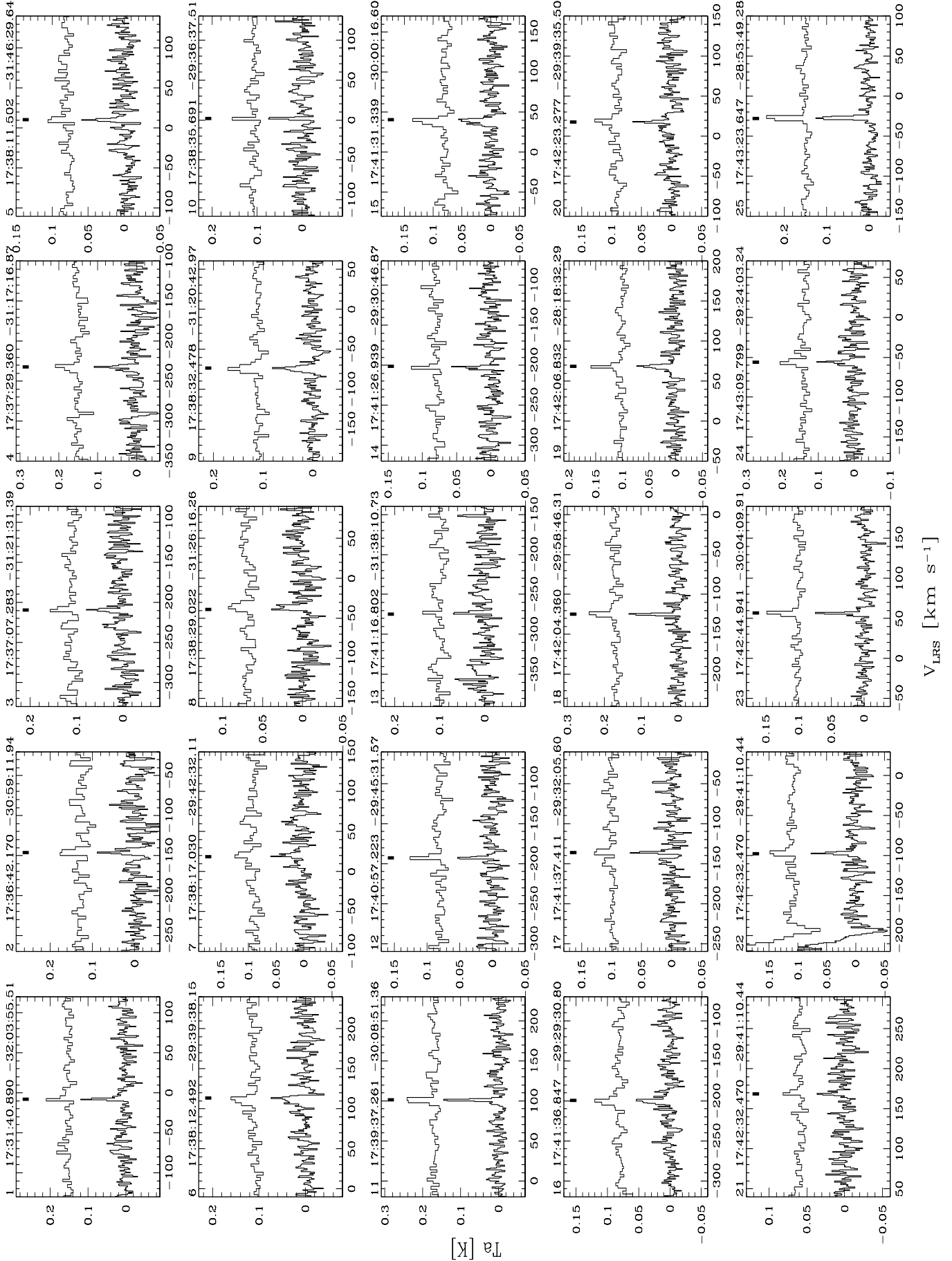}
\caption{\label{fig:spectra}
Spectra of SiO ($J = 2\rightarrow 1, v = 1$). Each panel 
shows the spectrum  obtained with the autocorrelator 
(lower spectrum) and  the one obtained with the filter
 bank (upper spectrum). 
The latter is shifted arbitrarily upwards for clarity.
The full figure shows 271 spectra and  is available
only in the electronic version of the paper at http://www.edpsciences.org.}
\end{figure*}

Figure  \ref{fig:spectra}  shows the  spectra of the detected 
SiO ($J = 2\rightarrow 1, v = 1$) lines. In each panel  the 
spectrum  obtained with the autocorrelator (lower spectrum) and  
the one obtained with the filter bank (upper spectrum) is given. 
The latter is shifted arbitrarily upwards for clarity. 

The Table with all the additional IR measurements 
(from DENIS-ISOGAL and from MSX data) will follow in the 
next paper where the physical properties of 
the sources will be discussed \citep[][ in preparation]{messineo02b}.

\begin{table}[h]
\caption{\label{table: non-detections} Sources with no SiO maser 
detected$^{\mathrm{*}}$.The conversion factor from antenna 
temperature to flux density is 6.2 Jy K$^{-1}$.} 
\medskip  
\ \\ 
$$ 
\begin{array}{ccccll} 
\noalign{\smallskip}  
\noalign{\hrule} 
\noalign{\smallskip}  
 {\rm ID} & {\rm RA }     & {\rm DEC}  &  {\rm rms}  &  {\rm Obs. Date} & {\rm Comments}  \\ 
           &  [{\rm J}2000]&  [{\rm J}2000] & [{\rm K}] & [{\rm yymmdd}]       &         \\ 
\noalign{\smallskip}  
\noalign{\hrule} 
  272   & 17\:31\:57.6  & $$-$$32\:14\:11 &  0.018  &       010216 &             \\
  273   & 17\:35\:56.2  & $$-$$31\:40\:42 &  0.016  &       010528 &             \\
  274   & 17\:37\:04.2  & $$-$$27\:52\:04 &  0.017  &       010523 &             \\
  275   & 17\:37\:07.7  & $$-$$27\:51\:06 &  0.016  &       010523 &             \\
  276   & 17\:37\:42.9  & $$-$$31\:24\:56 &  0.021  &       010216 &             \\
  277   & 17\:38\:01.6  & $$-$$29\:46\:60 &  0.010  &       000827 &             \\
  278   & 17\:38\:26.5  & $$-$$31\:28\:22 &  0.011  &       010815 &             \\
  279   & 17\:39\:29.8  & $$-$$30\:10\:20 &  0.011  & 010219/010903 &             \\
  280   & 17\:39\:30.4  & $$-$$30\:13\:33 &  0.012  &       000826 &   41172$$^{\mathrm{b}}$$    \\
  281   & 17\:39\:30.7  & $$-$$30\:08\:50 &  0.016  &       010219 &   37877$$^{\mathrm{b}}$$    \\
  282   & 17\:39\:35.7  & $$-$$31\:53\:42 &  0.019  &       010523 &             \\
\noalign{\smallskip}  
\noalign{\hrule} 
\end{array} 
$$ 
\begin{list}{}{}
\item[$^{\mathrm{*}}$] 
The full table contains 173 objects and is available in electronic 
form at the CDS via anonymous ftp to ftp://cdsarc.u-strasbg.fr
or via http://cdsweb.u-strasbg.fr/cgi-bin/qcat?J/A+A/(vol)/(page).
\item[$^{\mathrm{a}}$] 
Identification number from Table 2 of \citet{glass01}.
\item[$^{\mathrm{b}}$]
Identification number from Table 2 of \citet{schultheis00}.
\end{list} 
\end{table} 

\subsection{Detection rate}

We have observed 441 positions, and detected SiO ($v = 1, J = 2 \rightarrow
1$) maser lines in 268 of them. Since 3 spectra show two SiO lines at
different velocities ( \#21 and \#22; \#64 and \#65; \#77 and \#78)  
the total number of detected lines is 271  (see Table \ref{table:detections}
for the detections and \ref{table: non-detections} for non-detections).
The total  detection rate is $61 \%$. 

The spectra with two detections in one single beam are most probably  
detections  of another AGB masing star by chance in the beam (29\arcsec) 
of our targeted sources (see Sec.\ 4.6). 
The number of these detections is a function of the stellar density:
the three chance detections are located within 
$1^\circ$ from the Galactic centre.
Considering that chance detections  are distributed 
randomly among the overall detections and non-detections 
of targeted sources, we  deduce 6 as the number of 
chance detections within one degree from the Galactic centre. 
This corresponds to 5 \% of the 123 observations performed in that region, 
which in total cover 86 square arcmin. 
The obtained spatial density of chance detections is  consistent 
with the 43 GHz SiO maser density, 360 sources per square degree 
(8.5 sources in 86 square arcmin), obtained in  the Galactic centre  
by \citet{miyazaki01}.
This indicates that any blind survey will be less efficient than a targeted survey
even in the central few degrees of our Galaxy.

\begin{figure}[h]
\resizebox{\hsize}{!}{\includegraphics{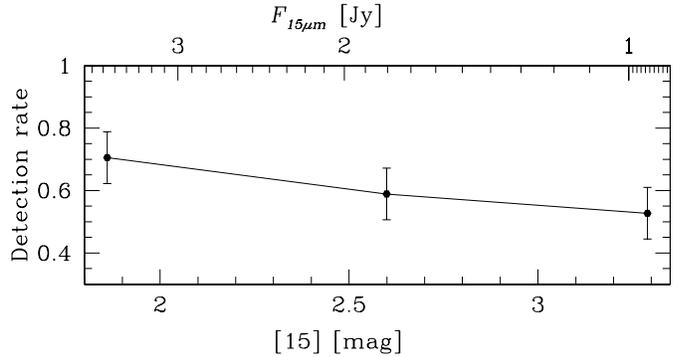}}
\caption{\label{fig:dr.ps} Detection rate  as a function
of the magnitude at 15 $\mu$m (ISOGAL $[15]$ or MSX D band).
Each bin contains $\sim$ 145 sources.
}
\end{figure}

The SiO maser detection rate tends to slightly 
increase with the mIR flux density at 7 and 15 $\mu$m.  
In Fig.\ \ref{fig:dr.ps} we show the detection rate as function 
of the (ISOGAL) magnitude at 15 $\mu$m, $[15]$, or the MSX D band magnitude
if no 15 $\mu$m ISOGAL magnitude is available.
The detection rate is 71 \% for the bright mIR sources at magnitude $[15] \sim$ 1.8 ($\sim 3.8$ Jy), and 
decreases to 53 \% for the less bright mIR sources at magnitude $[15] \sim$ 3.2 ($\sim 1$ Jy).

\begin{figure}[h]
\resizebox{\hsize}{!}{\includegraphics{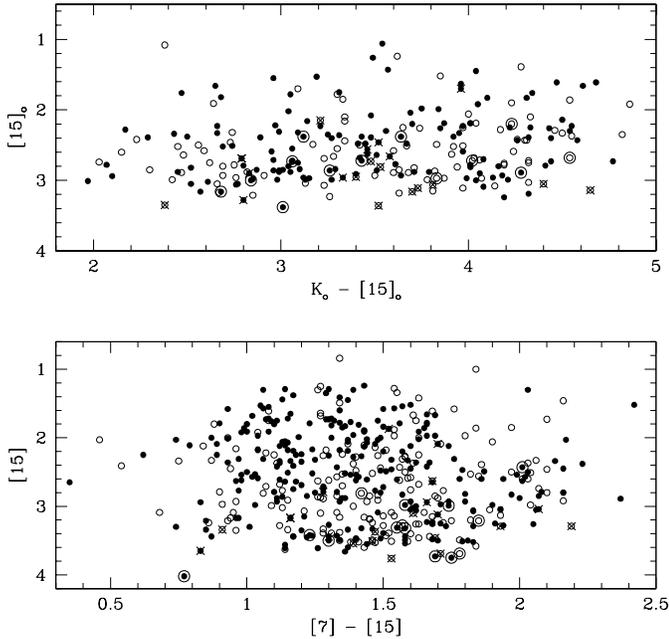}}
\caption{\label{fig:drcmd.ps} Lower panel: 
$[15]$ versus  $([7] - [15])$ colour-magnitude diagram; 
where ISOGAL magnitudes, or the A and D MSX magnitudes
if no  ISOGAL magnitude is available, are used. 
Upper panel:    $[15]_o$ versus $(K_o - [15]_o)$  ISOGAL-DENIS 
extinction corrected colour-magnitude diagram.
Detections are shown as filled circles; non detections as
open circles. The LPVs from \citet{glass01}
are marked with a larger open circle, whereas the
sources from \citet{schultheis00} are marked with crosses.
Note that two points  fall outside the figure.
 }
\end{figure}

We detected SiO maser emission in 143 out of 253 targets observed 
from the ISOGAL catalogue (57 \%). 
The MSX targets give a higher
detection rate: 125 detections out of 188 sources (66 \%). 
This is due to the correlation between the mIR flux density and the detection rate
and to the different  sensitivity of the ISOGAL and MSX surveys.
In fact, the ISOGAL sources were selected to have
magnitude at 15 $\mu$m lower than 3.4, i.e., flux density larger than $\sim$800 mJy,
while most of the MSX targets have a  flux density in the
 D band  higher than 1.5 Jy.  
If we restrict our analysis to the brightest ISOGAL sources
( $F_{15} > 1.5$ Jy, $[15] < 2.75$), we find similar results for 
both samples.

\subsection{Variability}
We observed  15 LPVs  found by \cite{glass01} and 
detected  SiO maser emission from 11 of them (73 \%). 
Since the observations were taken at a random pulsation phase
and since the SiO maser intensity is known to vary 
during the stellar phase by up to a factor ten \citep{bujarrabal94},
this detection rate is a lower limit to the actual 
percentage of LPV sources characterized by 86 GHz SiO maser emission.
Only 16 \% of those LPV stars have associated 
OH emission \citep{glass01}, and among those within our  defined
colour-magnitude regions  only 23 \%.   
Among large amplitude variable AGB stars the 86 GHz SiO masers 
are much more frequent  than OH masers.

Our sample of ISOGAL-DENIS sources also includes 19 sources from a list of 
\cite{schultheis00} of candidate variable stars, which were selected on the basis 
of repeated observations within the DENIS survey. We detected SiO maser emission
in 8 of those stars. The low detection rate in these
candidate variable stars may be due to  their 
low  mIR brightness,
$[15] \sim$ 3.2 (see Fig.\ \ref{fig:drcmd.ps} and  \ref{fig:dr.ps}),
and the uncertain indication of variability.

For the rest of our sources the only available information on 
variability is given by the photometric flag in the MSX catalogue
\citep{egan99}.
The sources we selected from the MSX catalogue all have an indication
of variability in band A. Of  the  ISOGAL-selected sources 
with a MSX counterpart \citep[][ in preparation]{messineo02b}, 
about half show variability in at least one  MSX band.
The ISOGAL catalogue does not contain any variability information.
\citet{alard01} have combined ISOGAL and
MACHO data in Baade's Windows and found   
that 90 \% of the objects detected in the MACHO and
ISOGAL  show well-defined variability (SR and Mira stars);
however, for most SRs the  amplitude of the variation is small.
The Mira stars among these are generally  the most luminous dust 
emitters \citep[Fig.\ 1 in][]{alard01}.
With $[15] < 3.4$, our sources  are  brighter than the Mira stars 
in Baade's Windows, the latter having shorter
periods than the Galactic centre LPVs  and  lower luminosity 
\citep{glass01,blum96}.
Thus, most of our sources are probably strongly variable long period AGB stars.
Follow-up  variability studies are  recommended.

\subsection{Line intensity}

In spite of the many observational studies of  SiO maser
emission, its pumping mechanism is still unclear.  
Previous 43 GHz SiO maser and mIR observations show a linear
correlation between the respective flux densities
\citep{bujarrabal87,nyman93,bujarrabal96,jiang02}. 
This correlation argues
 in favor of  radiative pump of the SiO masers, 
and  against collisional pumping models.
The average ratio between the 86 GHz 
 SiO ($v = 1$) maser peak intensity and the
12 $\mu$m IRAS flux density is 
0.1, though with a large scatter \citep{bujarrabal96}.

\begin{figure}
\resizebox{\hsize}{!}{\includegraphics{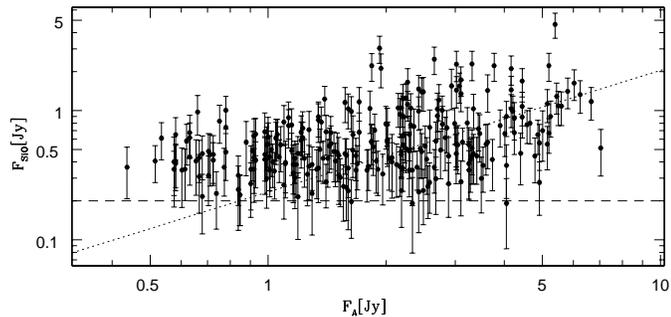}}
\caption{\label{fig:fig10.ps} 
86 GHz SiO  peak intensity as a function of
MSX A band or ISOGAL 7 $\mu$m flux density.
The dashed horizontal line
shows the average 3$\sigma$ detection limit. 
The dotted line shows the relation 
obtained by \citet{jiang02} from 43 GHz data.
}
\end{figure}

The dotted line in
 Fig.\ \ref{fig:fig10.ps} is the best fit found
by  \citet{jiang02} between the 43 GHz SiO maser intensity
and the MSX band A flux density.
Our results do not constrain the
 linear relation between the SiO maser and the mIR
flux densities.
Unfortunately, our data are not suitable to study
this relation because the SiO intensity distribution is 
limited by  sensitivity and the data  span  less than one order
of magnitude of the mIR flux density, 
which is  narrower than  the data of previous work.
The scatter is caused partly by the intrinsic source variability
 and the non-simultaneity of the mIR and SiO
maser observations, and partly by a wide range of  source distances.
We looked at the distance effects considering the magnitudes
log($F_{SIO}/F_D)$ and log($F_A/F_D)$, which are independent of
the distance, and we obtained a similar scattered diagram.

\subsection{ Longitude-velocity diagram}

\begin{figure}[h]
\resizebox{\hsize}{!}{\includegraphics[angle=-90]{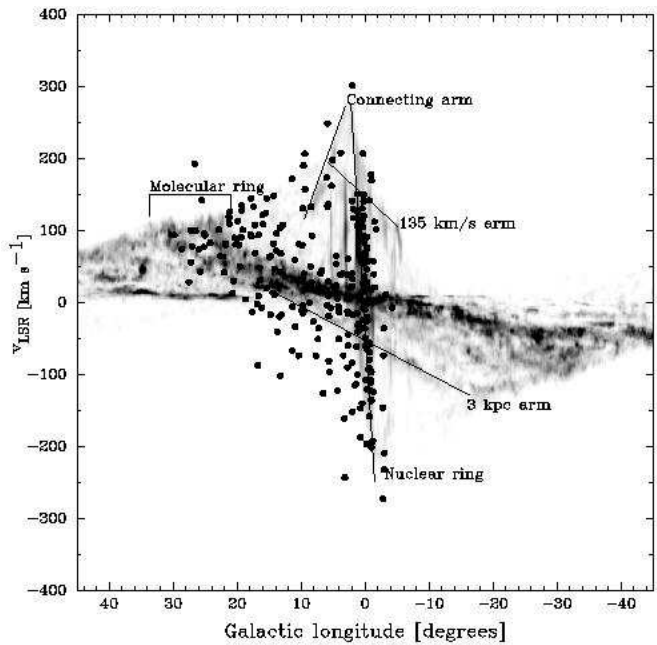}}
\caption{ \label{fig:vl86} 
Stellar longitude-velocity diagram overlayed on the grayscale CO ($l-v$)
diagram from \citet{dame01}.  The SiO 86 GHz masers are shown as dots.} 
\end{figure}
\noindent
In Figure  \ref{fig:vl86} we compare the longitude-velocity ($l-v$)
distribution of our SiO maser stars
with that of the $^{12}$CO line emission \citep{dame01}.  
The gas kinematics is a good tracer of
the dynamical mass in the Galaxy.  Main features of the 
distribution of the CO in the Inner Galaxy are labelled in Fig.\
\ref{fig:vl86}, following Fig.\ 1  of \citet{fux99}.  
The line-of-sight velocities of our SiO maser sources range 
from $-$274  to 300 \kms, consistent with previous stellar
maser measurements and with the CO velocities. 
An appreciable number of SiO sources are away from the CO emission
contours, in a region forbidden  for  pure circular rotation,
at negative velocities between $20\degr > l > 0\degr$. Around
longitude $0\degr$ the stellar distribution  follows the high
velocity gas component of the nuclear disk.
Similar results were found by \citet{sevenster01}.
We defer  a more detailed analysis of the 
kinematic properties of the new SiO maser sample to a future paper.  

\begin{figure}
\resizebox{\hsize}{!}{\includegraphics[angle=-90]{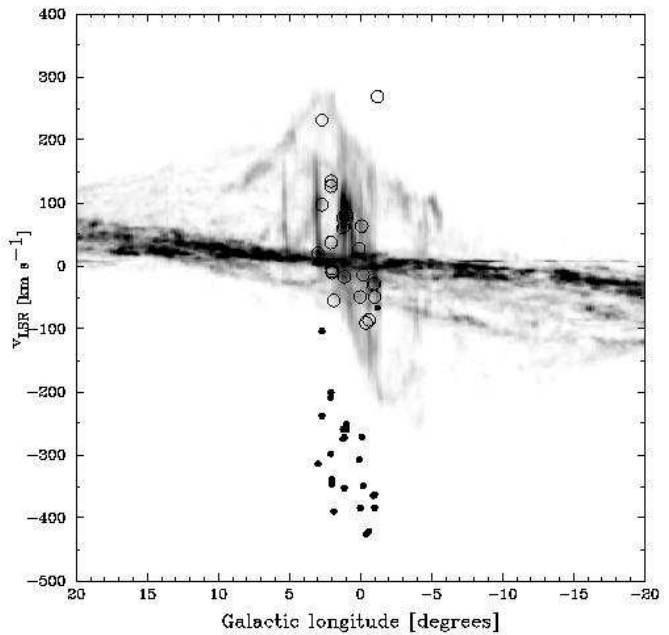}}
\caption{\label{figure:aa3.ps}
The grayscale  is the CO ($l-v$)
diagram from \citet{dame01}. The open circles indicate   
 \hcn\ lines,  their location is in agreement with the CO distribution.
The dots  represent the
position of the same lines if considered as SiO lines. The ($l-v$) diagram
confirms the interstellar origin of those lines. The open circle  
located above the CO distribution at $l=-1.2$ (source \#288) is discussed in Sec. 4.6.}

\end{figure}

In Figure \ref{figure:aa3.ps} we have marked, as open circles,
the  location of the \hcn\ lines we 
detected at velocities smaller than $-$30 \kms\  with 
widths wider than 7.5 \kms.
Their  distribution mostly follows the central gas  distribution,
confirming their interstellar origin and  the validity of the adopted
classification criteria based on the line width.

\subsection{Comparison with previous detections}

All OH/IR stars from the catalogues by \citet{sevenster97a,sevenster97b,
sevenster01,sjouwerman98a} and  \citet{lindqvist92}
were excluded from our target list since their
velocities are already known from the OH maser lines. Besides,
previous studies \citep{nyman86,nyman93}  anticipate a low
detectability of 86 GHz SiO masers in OH/IR stars toward the Galactic
centre,  which also seems to be consistent with the results of
our recent 30-m IRAM survey for 86 GHz SiO masers in Galactic center
OH/IR stars \citet{messineo02a}, in preparation.

\subsubsection{Nobeyama 43 GHz surveys}

There is a small overlap between the regions observed for 
the Japanese 43 GHz SiO maser surveys of IRAS 
point sources \citep{izumiura99,deguchi00a,deguchi00b} 
and  the region observed for our survey.
Using a search radius of 25\arcsec\ around each of our sources, 
which is about half the 43 GHz main beam at the Nobeyama telescope 
and almost the full 86 GHz beam of the IRAM 30-m, 
we found 19  matches between our positions and the IRAS sources
positions (Table \ref{table: deguchi}).
The velocities at 86 and 43 GHz of 7 sources detected
in both lines agree within a few \kms. 
Source \#256 (IRAS 18301-0900) shows a difference of 8 \kms\ 
between the SiO maser line at 43 GHz and at 86 GHz, but this
source  clearly has a double peak in the 43 GHz spectrum and the 86 GHz
peak corresponds to one of the 43 GHz peaks.  The low number of
sources which are detected in both surveys needs further study, but 
can partially be due to source variability. 
Considering those 19 sources in common between the two surveys,
the SiO maser detection rate appears higher at 43 GHz (68  \%)
than at 86 GHz (52 \%).  However, this small 
sample of  sources  is not
representative of our full sample and of our 86 GHz SiO maser detection rate
(66 \% for MSX); it is biased toward  redder A-D colour 
(higher mass-loss rate).

\begin{table} [h]
\caption{\label{table: deguchi} Overlap between our 86 GHz SiO maser survey and
the 43 GHz SiO maser survey  in IRAS sources by \citet{izumiura99,deguchi00a,deguchi00b}. 
Column (1) lists the source ID number, column (2) the IRAS name,
 column (3) the 86 GHz line-of-sight
velocity,  column (4)   the 43 GHz line-of-sight
velocity and column (5) the separation between  the IRAS position  
and our position. 
}
\medskip  
\ \\ 
$$ 
\begin{array}{rrrrrr} 
\noalign{\smallskip}  
\noalign{\hrule} 
\noalign{\smallskip}  
 {\rm ID}   & {\rm IRAS} & \multicolumn{2}{c}{\rm V$$_{\rm LSR}$$}    & {\rm separation} &  \\ 
       &    & \multicolumn{2}{c}{\rm [km\ s$$^{-1}$$]}                     & {\rm [arcsec]}   &  \\
       &    & 86 {\rm\ GHz}            &  43 {\rm\ GHz}                     &            &  \\ 
\noalign{\smallskip}  
\noalign{\hrule} 
             72    &  17429$$-$$2935    &     2.6     &     $$-$$0.8  & 22.2 & \\     
            125    &  17497$$-$$2607    &  $$-$$162.5 &   $$-$$163.1  &  7.6 & \\     
            141    &  17524$$-$$2419    &     196.0   &  {\rm  no\ det}& 10.8 & \\     
            154    &  17563$$-$$2402    &     136.2   &  {\rm  no\ det}& 15.9 & \\     
            162    &  17586$$-$$2329    &  $$-$$125.6 &  {\rm  no\ det}&  7.6 & \\     
            166    &  18000$$-$$2127    &     134.1  &    134.1      &  6.1 & \\     
            187    &  18056$$-$$1923    &   $$-$$34.3 &    $$-$$32.2  &  8.7 & \\     
            197    &  18106$$-$$1734    &   $$-$$16.9 &    $$-$$16.7  & 21.6 & \\     
            236    &  18229$$-$$1122    &     112.3   &    111.4      &  4.5 & \\     
            256    &  18301$$-$$0900    &     101.5   &    109.6      &  5.5 & \\     
            361    &  17483$$-$$2613    &{\rm  no\ det}&    163.2      &  7.0 & \\     
            363    &  17492$$-$$2636    &{\rm  no\ det}&     88.8      & 14.9 & \\     
            365    &  17497$$-$$2608    &{\rm  no\ det}&  {\rm  no\ det}&  4.7 & \\     
            371    &  17509$$-$$2516    &{\rm  no\ det}&    $$-$$93.8  &  5.3 & \\     
            400    &  18037$$-$$2026    &{\rm  no\ det}&     68.2      &  6.4 & \\     
            411    &  18167$$-$$1517    &{\rm  no\ det}&  {\rm  no\ det}& 12.3 & \\     
            417    &  18238$$-$$1135    &{\rm  no\ det}&    180.0      &  6.9 & \\     
            418    &  18246$$-$$1125    &{\rm  no\ det}&     80.0      &  2.8 & \\     
            423    &  18302$$-$$0848    &{\rm  no\ det}&  {\rm  no\ det}&  9.2 & \\     
\noalign{\smallskip}  
\noalign{\hrule} 
\end{array} 
$$ 
\end{table} 

To avoid saturation of the detector, the very centre of the Galaxy was 
not observed by ISOGAL \citep[see][]{ortiz02}.
Therefore, none of our sources is located in the regions centered 
on SgrA, which were mapped at 43 GHz by  \citet{miyazaki01} and 
\citet{deguchi02}. 

\citet{imai02} report   43 GHz SiO maser detections 
towards  LPV stars found by \citet{glass01}. 
Seven LPV stars detected at 43 GHz  
coincide with sources in our 86 GHz SiO maser survey
(\#48 = g23$-$5, \#49 = g23$-$8, \#52 = g21$-$39, \#56 = g22$-$11,        
\#73=g6$-$25, \#80=g14$-$2, \#332=g12$-$21).
In 4 cases there is a corresponding SiO maser detection in the 86 
GHz spectrum at  a velocity consistent with the 43 GHz line. 
Sources \#80 and \#49 were both detected at 86 GHz, but 
there is a significant difference between the 86 and 43 GHz velocities
of 7 and 17 \kms, respectively. 
The reason for this is unclear because the 43 
GHz spectra were not published in Imai et al (2002).
Finally,  \#332 was detected at 43 but not at 86 GHz,
which is probably due to source variability.

\subsubsection{43 GHz SiO masers in the Sagittarius B2 Region}

\begin{table}[h] 
\caption{\label{table:sgrb2} 
Cross identification with sources detected by \citet{shiki97}.  
Reference numbers from \citet{shiki97} 
and our names are listed in column 1,
and 2 respectively.  The separation from 
the closest ISOGAL source and its name in column 3 and 4. Finally
the velocities at 43 and 86 GHz in columns 8 and 9. 
Shiki's source number 2 is not in the ISOGAL catalogue
prepared for the first release, but it is clearly a strong 
source in the ISOCAL image, 
and probably is associated with a foreground supergiant 
(Shiki et al 1997).
 } 

\medskip  
\ \\ 
$$ 
\begin{array}{rrrrrr} 
\noalign{\smallskip}  
\noalign{\hrule} 
\noalign{\smallskip}  
{\rm REF}&{\rm ID} & {\rm separation} &{\rm ISOGAL\ name}&   \multicolumn{2}{c}{\rm V$$_{\rm LRS}} \\
         &         &    {\rm [arcsec]}    &                    & \multicolumn{2}{c}{{\rm [km\thinspace s}$$^{-1}$$]}    \\
         &         &                &                  &   {\rm 43\ GHz} & {\rm 86\ GHz}    \\
\noalign{\hrule} 
1&88 &   6.51 & PJ174656.2$$-$$283105 &        56.5& 55.8      \\
2&   &        &                       &   $$-$$25.7&           \\  
3&   &   2.03 & PJ174720.3$$-$$282305 &        82.3&           \\  
4&   &   7.19 & PJ174812.0$$-$$281817 &        80.9&           \\  
5&   &  16.46 & PJ174823.7$$-$$282018 &        81.0&           \\  
6&   &  10.56 & PJ174814.6$$-$$280852 &       101.2&           \\  
7&99 &   6.95 & PJ174813.2$$-$$281941 &   $$-$$38.7& $$-$$36.5\\
\noalign{\smallskip}  
\noalign{\hrule} 
\end{array} 
$$ 
\end{table} 
\citet{shiki97} mapped a subregion of Sgr B2 at 43 GHz
and found seven SiO maser lines, only two of which  
could be identified with an IRAS source (Shiki's \#3 and \#6).
We found an ISOGAL counterpart within 17\arcsec\ in  all cases, and from their
mIR colours, we confirm that those seven detections are 
stellar maser sources in the  line-of-sight of the  giant  molecular cloud Sgr B2.
Two of those ISOGAL sources were observed in our 86 GHz survey, and 
for both sources  the 43 and 86 GHz line velocities are consistent 
(see Table  \ref{table:sgrb2}).

\subsection{Comments on individual objects}

\noindent
{\bf \#21 and \#22}\\
Sources \#21 and \#22 were detected in the same beam.  The peak
intensity of \#21 is only 3 times the noise rms and we list the line as a
marginal detection.  However, a second ISOGAL source,
ISOGAL$-$PJ174232.9$-$294124, happened to fall inside the beam, at 
15.6\arcsec\ from the  position we targeted.  This source is less bright at 15
$\mu$m
($[15] = 4.73$) than the targeted ISOGAL$-$PJ174232.5$-$294110
($[15] = 3.17$) and this suggests that the original targeted
ISOGAL$-$PJ174232.5$-$294110 is the
mIR  counterpart of the stronger SiO line, \#22, while
ISOGAL$-$PJ$174232.9-294124$ is probably the counterpart of  \#21.
Observations at 86 GHz and/or both of the 43 GHz SiO lines, at both stellar
positions, may confirm our conclusion.

\bigskip
\noindent
{\bf \#64 and \#65}\\
Two very narrow SiO line sources \#64 and \#65 were detected in the
same spectrum.  Both
lines have the same peak intensity, but different velocities, 53.7
and $-$7.2 \kms, respectively. This
suggests that the two lines are generated in  the envelopes of two different AGB stars.
The ISOGAL
catalogue does not give  another source within 30\arcsec\
of the position of targeted ISOGAL-PJ174528.8-284734, neither does
inspection of the ISOGAL images at 7 and 15 $\mu$m. However, at that
position there is
strong background emission which may have limited detection of
fainter stellar mIR sources.

\bigskip
\noindent
{\bf  \#77 and \#78}\\
Sources \#77 and \#78 were 
detected in the same spectrum at velocities
of 141.6 and 27.6 \kms, respectively.  The  targeted ISOGAL source,
ISOGAL-PJ174618.9-284439, is separated by  12.5\arcsec\ from
ISOGAL-PJ174619.5-284448. However, the latter is a weak mIR source
only detected  at 7 $\mu{\rm m}\ ([7] = 8.73)$ by ISOGAL.
Again, observing at 86 or 43 GHz at both stellar positions
may resolve the mIR counterpart associated with the SiO maser line.

\bigskip
\noindent
{\bf \#94}\\
This double peaked source is located in the Sgr B2 region.
The two lines have similar intensities and are at velocities  
$-36.6$ and $-$28.8 \kms\ with respect to SiO (or 299.4 and 307.2 with respect to \hcn),
with  widths of 5 and 3 \kms, respectively. The small velocity separation between
the peaks suggests that the two emissions are related.
The  velocity separation is also consistent with  two different  
\hcn\ hyperfine transitions, but one of the two peaks has a velocity larger than $-30$
\kms\ and does not fall in our  \hcn\
classification criteria. The location of this source on the ($l-v$)
diagram agrees with the CO distribution when considered
as an SiO line. Thus the source is listed here among the  
SiO line detections.
In Sgr B2, other double peaked profiles have been seen in SiO emission
with line widths of $\sim 100$ \kms\ and at velocities
from $\sim -$25 to $\sim 100$ \kms\ \citep{martinpintado97}.
The SiO emission in \#94  may  not  be  associated
with the circumstellar envelope close to the star as in all 
other cases, as its profile  may be more typical to that 
of bipolar molecular outflows.

\bigskip
\noindent
{\bf \#117}\\
See the discussion in Sect.\ 3.4.

\bigskip
\noindent
{\bf \#288}\\ 
We detected a 14.7 \kms\ wide line at a velocity of $-$66.6 \kms\  with 
respect to  SiO (or 269.4 with respect to \hcn),  which we classified as likely being an \hcn\ line.
However, Fig.\ \ref{figure:aa3.ps}\ shows  that the point if  regarded as \hcn\
is far from any  CO emission.
The source, $(l, b)=(358.779^\circ, 0.227^\circ)$,
is located in the region
of the X-ray transient (EXS17379$-$2952), a region of interest 
to many other observers.
In that region, \citet{durouchoux98} detect a few dense CO molecular clouds, of which one at 
a velocity of $-60$ \kms. 
We suggest that the SiO line  at the position of \#288
has an  interstellar origin  and is associated with the CO cloud of 
\citet{durouchoux98}.

\bigskip
\noindent
{\bf \#423}\\
At the position of \#423 we detected a line which according to our
criteria is an \hcn\  line. This is the only detection outside the
Galactic centre region, at a longitude of 23 degrees,
that is found at high negative velocity, $-204.7$ \kms\ (with respect to SiO)
and with a fairly wide line width (22 \kms).
Its position as SiO line does not fit the velocity-longitude diagram (it
does fit when regarded as \hcn\ emission, then at velocity 131.3 \kms), and
\citet{izumiura99} searched for 43 GHz SiO maser without any success.
The MSX maps do not show any extended mIR emission or dark region  at that
position that could suggest the presence of a cloud, however CO maps
show a strong concentration of molecular matter \citep{dame01}.
Also IRAS detected a mIR source, IRAS18302$-$0848, within 10\arcsec\
from the MSX position, and with IRAS  flux densities consistent with the
MSX flux densities.
For this highly reddened source, \citet{stephenson} found a strong excess
(4-5 magnitudes) in the $R-I$ colour and absence of  molecular bands in the
$I$-spectrum, and concluded that any intrinsic contribution to the redness
should be small.

Following his conclusion, that IRAS 18302-0848 is a distant luminous star
\citep[which has also been supported by][]{creese95},
we conclude that \#423 is not an 86 GHz SiO maser emitter and that
the origin of reddening of this star is also the origin of the \hcn\ 
emission we detected.

\section{Conclusions}
We have observed 441 colour-selected ISOGAL and MSX sources in the Inner
Galaxy ($30^\circ < l < -4^\circ$ and $|b|$ mostly $<1$), in the SiO
($v = 1, J = 2 \rightarrow 1$) maser transition and  detected 271  lines.
We thereby obtained 255 new line-of-sight velocities which doubles the number
 of maser line-of-sight velocities  known in  the region we surveyed.
To search for 86 GHz ($v = 1$) SiO maser lines in colour-selected mIR sources 
has proven to be an efficient way to  obtain stellar radial velocities in the Inner Galaxy.
In the central 2 degrees we notice some confusion with interstellar \hcn\ emission, 
but usually the interstellar \hcn\  and the stellar SiO line 
can be separated  well  by using their radial velocities and line widths. 
The SiO maser emission was detected towards  61 \%  of our 
sources, objects which lie in a transition region of the IR-colour space between
Mira and OH/IR stars.
The SiO maser detectability decreases with decreasing mIR flux density. 
We observed 15 sources from the sample of LPV stars by \citet{glass01}
and found  86 GHz SiO maser emission in 11 of them (73 \%), 
while only 23 \% of the LPV stars which follow our selection criteria
show OH maser emission. 
Therefore 86 GHz SiO maser emission is more 
frequent than OH maser emission.
In a later study we will use our new catalogue of 
stellar line-of-sight velocities
for a quantitative analysis of stellar kinematics 
and SiO maser properties  in the Inner Galaxy.

\begin{acknowledgements}  
 We thank D. Levine and M. Morris for sharing their experience 
about preliminary 
observations of SiO masers with the IRAM 30m telescope.
We are grateful to Ute Lisenfeld, Frank Bertoldi, and the IRAM
staff for their support in the observations, most of which
were made possible only through the flexible observing
mode recently introduced at the 30m telescope.
We thank Frederic Schuller for his help with the ISOGAL data.
Many thanks to Martin Bureau for fruitful discussions on stellar  
galactic dynamics.  This work was carried out in the context of EARA, 
the European Association  for Research in Astronomy. 
LOS acknowledges support from the European Commission under
contract HPRI-CT-1999-00045. The work of MM is funded by the Netherlands
Research School for Astronomy (NOVA) through a {\it
netwerk 2, Ph.D. stipend}.
\end{acknowledgements}

\bibliographystyle{aa}
\bibliography{2744.biblio}

\end{document}